\documentclass[11pt]{article}
\usepackage{graphicx}
\usepackage[numbers, comma,sort&compress]{natbib}
\usepackage{amssymb}
\usepackage{amsmath}
\usepackage{array}
\usepackage{verbatim}

\newcommand{\thetitle}{Ponderomotive light squeezing with atomic cavity optomechanics}
\newcommand{\Dpc}{\Delta}
\newcommand{\nbar}{\bar{n}}
\newcommand{\Gm}{\Gamma_m}
\newcommand{\Go}{\Gamma_{\mathrm{opt}}}
\newcommand{\Gt}{\Gamma_{\mathrm{tot}}}
\newcommand{\wm}{\omega_m}
\newcommand{\ws}{\omega_s}
\newcommand{\wc}{\omega_0}
\newcommand{\sopt}{s_{\mathrm{opt}}}
\newcommand{\aop}{\hat{a}}
\newcommand{\bop}{\hat{b}}

\newcommand{\afop}{\tilde{a}}

\newcommand{\bfop}{\tilde{\zeta}}

\newcommand{\alfop}{\tilde{\alpha}}
\newcommand{\fb}{\mathrm{f_{beat}}}
\newcommand{\gom}{g}
\newcommand{\zzp}{z_{\mathrm{zp}}}
\newcommand{\dcaption}[1]{\caption{\baselineskip 18pt #1}}

\begin{document}

\baselineskip 18pt

\title{\textbf{\thetitle}}
\author{
     Daniel W. C.\ Brooks$^1$\thanks{Email: dwb@berkeley.edu}, Thierry Botter$^1$, Nathan Brahms$^1$, Thomas P.\ Purdy$^{1}$\thanks{Present address:\ JILA, University of Colorado, Boulder, CO 80309, USA},\\ Sydney Schreppler$^1$, and Dan M.\ Stamper-Kurn$^{1,2}$
   \\ \\
  \small{1. Department of Physics, University of California, Berkeley, CA 94720, USA}\\
  \small{2. Materials Sciences Division, Lawrence Berkeley National Laboratory,}\\
 \small{Berkeley, CA 94720, USA}
}
\date{\today}
\maketitle
\thispagestyle{plain}
\begin{abstract}
Accessing distinctly quantum aspects of the interaction between light and the position of a mechanical object has been an outstanding challenge to cavity-optomechanical systems. Cold-atom implementations of cavity optomechanics have demonstrated sensitivity to the quantum fluctuations in the optical radiation pressure force. Here we implement such a system, in which quantum photon-number fluctuations significantly drive the center of mass of an atomic ensemble inside a Fabry-P\'{e}rot cavity. We observe sub-shot-noise ponderomotive squeezing of quantum fluctuations and also demonstrate that classical optical fluctuations can be attenuated by 26~dB or amplified by 20~dB with a weak input pump power of $<40$~pW. The observation of squeezing and characterization of the optomechanical amplifier's frequency-dependent gain and phase response opens a route to enhancing force sensing with cavity-optomechanical systems.
\end{abstract}

The desire to measure ever weaker forces has spurred the development of cavity-optomechanical systems \cite{Kippenberg2008}, in which mechanical motion as minute as attometers \cite{Abbott2009} is converted into detectable optical signals. The very act of measurement requires a corresponding disturbance of the object due to the radiation-pressure force exerted by the light field upon the mechanical element \cite{Braginskii1967, Caves1980, Meystre1985}. In cavity optomechanics, this disturbance feeds back onto the light field, resulting in high-gain nonlinear parametric amplification of the electromagnetic field.  Myriad classical parametric effects arise, including motional damping \cite{Braginskii1970}, bistability \cite{Dorsel1983, Gozzini1985, Gupta2007}, induced transparency \cite{Teufel2011, Weis2010, Safavi-Naeini2011a}, and mode splitting \cite{Groblacher2009,Teufel2011}.  These effects have led to powerful applications of optomechanics, including regenerative amplification\cite{Rokhsari2005}, optical filtering \cite{Eichenfield2007}, cavity-sideband cooling to the motional ground state \cite{Teufel2011a,Chan2011}, and proposals to transduce disparate electromagnetic frequencies \cite{Safavi-Naeini2011, Regal2011} or forms of quantum information \cite{Stannigel2011}.

Here we achieve a long-standing goal of cavity-optomechanics research: the direct observation of an optomechanical parametric amplifier driven by quantum fluctuations. ``Ponderomotive squeezing'' is predicted to arise when a quantum-driven optomechanical amplifier causes frequency-dependent gain and attenuation of the optical spectrum \cite{Mancini1994, Fabre1994}.  Understanding and applying this squeezing will be critical for surpassing standard quantum limits on position and force sensing \cite{Pace1993,Clerk2010}, and for constructing the next generation of gravitational-wave observatories \cite{Kimble2001, Corbitt2006}. To date, thermal effects of the mechanical bath and optical technical noise have prevented the direct observation of a shot-noise-driven cavity-optomechanical system.  Therefore, experiments to date have focused on simulating ponderomotive squeezing by applying classical optical drives \cite{Marino2010, Verlot2010}.

In this work, we observe sub-shot-noise ponderomotive squeezing by realizing an atomic cavity-optomechanics system \cite{Gupta2007,Brennecke2008} that is driven primarily by quantum fluctuations. Our system consists of a cloud of ultracold atoms trapped in an optical lattice, and prepared near their motional ground state.  The center-of-mass motion of the atoms is linearly coupled to light, detuned many gigahertz from atomic resonance, in a high-finesse optical cavity.  Applying classical amplitude noise to the optical field, we first demonstrate that the system behaves as a high-gain parametric optomechanical amplifier with input pump power as low as 36~pW.  We then drive the system with near-shot-noise-limited light, and map the spectrum of ponderomotive squeezing \textit{vs.} frequency and detection quadrature.

The atomic ensemble is positioned at locations with a strong gradient of the probe intensity, yielding a linear optomechanical coupling between its position and the cavity resonance frequency. This coupling is encapsulated by the optomechanical Hamiltonian $H = \hbar \gom \aop^\dag \aop (\bop + \bop^\dag)$, where $\aop\: (\bop)$ and $\aop^\dag \: (\bop^\dag)$ are the annihilation and creation operators for the light (oscillator) field, and $\gom$ is the optomechanical coupling rate. For small fluctuations, the probe intensity can be linearized about the mean optical field. As a result, the cavity optomechanical system acts as a phase-sensitive linear amplifier for optical fluctuations \cite{Botter2011}.  The steady-state dynamical response of this amplifier at frequency $\omega$, neglecting additional external forces on the mechanical oscillator, is given by the relation between fluctuations of the intracavity field $\afop$ and  those of the input field $\bfop$ (after accounting for the cavity induced filtering; see Supplemental Information):
\begin{gather}
\label{Ha} \afop(\omega) = \bfop(\omega) + G(\omega)\,\frac{1}{2}\left[\bfop(\omega) + \bfop^\dag(-\omega)\right]\\
\label{Gw} G(\omega) \approx \frac{\left(i \kappa - \Dpc \right)\sopt}{\ws^2-\omega^2-i\omega \Gt}
\end{gather}
Amplitude fluctuations of the cavity-filtered input field, proportional to $\frac{1}{2}\left[\bfop(\omega) +\bfop^\dag(-\omega)\right]$, induce motion of the mechanical oscillator, which is then transduced back onto the cavity field with a closed-loop gain $G(\omega)$. In the above, $\kappa$ is the cavity half-linewidth, $\Delta$ is the probe detuning from cavity-resonance, and $\Gt = \Gm + \Go$ is the combination of mechanical and optomechanical damping.  The optomechanical stiffening parameter $s_\mathrm{opt} \propto \langle \aop^\dag \aop\rangle g^2$, shifts the mechanical resonance frequency from its unperturbed value $\omega_m$ to $\omega_s=\sqrt{\omega_m^2+\Delta s_\mathrm{opt}}$. Notably, the cavity field modulations induced by optomechanics can interfere destructively with the input fluctuations.  When technical and thermal fluctuations are sufficiently small, the ponderomotive attenuation of quantum fluctuations is resolved as sub-shot-noise squeezing in the power spectral density at frequencies near $\wm$.

We study the optomechanical response to both classical and quantum optical drives. For both experiments, we begin each run by using a microfabricated atom chip to place a gas-phase mechanical oscillator, composed of ultracold $^{87}$Rb atoms, within the Fabry-P\'{e}rot cavity.  After a stage of evaporative cooling, about 3500 atoms remain trapped within a few adjacent wells of the optical lattice.  The ensemble's center-of-mass motion along the cavity axis is thus prepared near its ground state and serves as the mechanical element \cite{Purdy2010}.  By controlling the intensity of the optical lattice, $\wm$ can be controlled over a range of frequencies. For these experiments, $\wm = 2\pi \times 155.5$~kHz is set significantly lower than the cavity half-linewidth $\kappa=2\pi\times1.8$~MHz.  The optomechanical coupling rate is then $\gom = \left( d \wc/dz \right) \zzp = 2\pi\times68$~kHz, where $\zzp = \sqrt{\hbar/2 m \wm}$ is the mechanical harmonic oscillator length, $\hat{z}=\zzp (\bop + \bop^\dag)$, the resonance frequency of the cavity is $\wc$, and $m=0.5$~attograms is the total mass of the atomic ensemble.

\begin{figure}[p]
	\centering
	\includegraphics[width=1\linewidth]{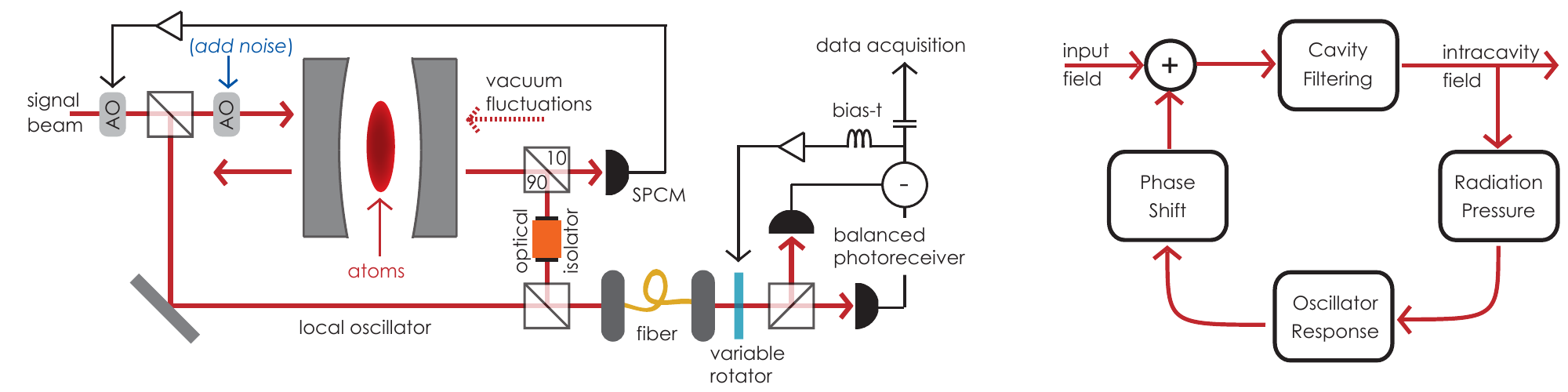}
	\dcaption{\textbf{(Left)} Schematic of cavity and heterodyne detection setup.  10\% of the cavity output light is diverted to a single-photon counting module (SPCM) using a polarizing beamsplitter (PBS), while the remainder goes to the heterodyne detection system. The SPCM signal is fed back to control the frequency of an acousto-optic modulator (AO), maintaining a constant detuning from the shifted cavity resonance. Additional AOs shift the signal beam 10 MHz relative to the local oscillator (LO) and can be used to add AM noise. The signal and LO are mode-matched through a fiber, and sent onto the detector, which is balanced by controlling beam polarization with a liquid-crystal rotator. \textbf{(Right)} Schematic of the cavity as an optomechanical amplifier with feedback. The intracavity field originates from input fluctuations that are filtered by the cavity and transduced by the optomechanical interaction.
}
	\label{fig:Schematic}
\end{figure}

As the optical amplification arises when the probe light is detuned from cavity resonance, we stabilize the probe frequency to $\Dpc = -1.0$~MHz. The light transmitted through the cavity is detected using a balanced heterodyne receiver, with an overall quantum efficiency of 10.1\% for detecting intracavity photons (Fig.~\ref{fig:Schematic}). To maximize the detection efficiency, the cavity mirror on the detector side is a factor of 8 more transmissive than the input mirror. We restrict the heterodyne signal record to the first 5 ms of probing, after which the atomic sample is significantly heated by radiation pressure fluctuations (see Supplemental Information).  In each experimental cycle, we also record two additional data streams from the heterodyne receiver: once after the optical cavity is emptied of atoms, and once with the probe light extinguished. These record the level of technical noise on the probe light and of shot-noise on the detector, respectively.

\begin{figure}[p]
	\centering
	\includegraphics[width=.7\linewidth]{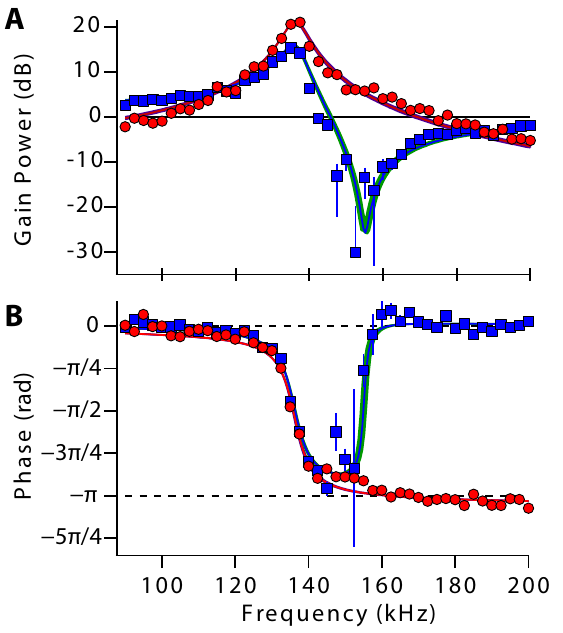}
	\dcaption{Optomechanical transduction of an classical AM drive. By fitting predicted responses to the data we extract the optomechanically shifted mechanical oscillation frequency ($\ws = 2\pi \times 136.0$~kHz) and the mechanical damping rate ($\Gm = 2\pi \times 1.91$~kHz).  \textbf{(A)} Square magnitude of the gain. The amplifier coherently amplifies the AM drive (blue squares) such that fluctuations are maximally squashed at $\wm$ by $-26$~dB below the response without amplification (black). The amplifier also transfers the drive into the PM quadrature (red circles), yielding a strong response centered at $\ws$. The data are in good agreement with the predicted AM (blue line) and PM (red line) gain power. The effect of systematic uncertainties on the theory are shown in the green and purple shaded regions. \textbf{(B)} Phase offset relative to drive tone. As a damped system at low frequencies, the amplifier adds a delay. At frequencies above $\wm$ , the optomechanical damping $\Go$ changes sign. When $-\Go > \Gm$ the phase response changes from delayed to advanced, crossing $0^\circ$ (-$180^\circ$) for the AM (PM) response.}
	\label{fig:Squash}
\end{figure}

 In the first experiment, following recent works \cite{Marino2010,Verlot2010}, we applied a strong amplitude modulation (AM) to the input optical field to measure the classical response of the optomechanical amplifier. The complex gain $G(\omega)$ produces amplitude and phase modulation (PM) in orthogonal quadratures of the intracavity field.  For each run, a single AM tone was applied.  Using a superheterodyne technique, we determined the complex response in both quadratures as a function of frequency. The response is characterized  (Fig.\ \ref{fig:Squash}) by a noise-power gain and phase difference relative to the input drive tone, which is independently measured by the empty-cavity data record.

The measured power gain shows regions of strong enhancement of fluctuations (up to 20~dB in power) at frequencies below $\wm$.  Remarkably, the large gain is achieved with only 36~pW of optical pump entering the cavity,  maintaining an average intracavity photon number of around 6.  This observation suggests that an optomechanical system could be applied in ultra-low-power photonics as an amplifier, filter or switch.  At higher frequencies, we observe strong suppression  (``squashing'') of amplitude fluctuations, with the maximal suppression of 26~dB at $\wm$.  The phase difference shows a transition from a delayed to an advanced response, indicated by the response crossing $0^\circ (-180^\circ)$ in the AM (PM) quadrature. The amplifier's stability is maintained since the gain is below unity at the frequencies having an advanced response. 

Having characterized the ponderomotive optical amplification of our system, we extinguish the deliberate AM tone and drive our system with a shot-noise dominated input field of nearly the same input power as before.  Under these conditions, fluctuations in the intracavity radiation-pressure force arise predominately from the vacuum field for frequencies near $\wm$.
Ponderomotive attenuation should result in a strongly squeezed optical field within the cavity with maximal squeezing of $\sim-20$~dB at $\wm$ in the AM quadrature (see Supplemental Information). Outside the cavity, interference with reflected shot noise shifts the optimally squeezed frequency and quadrature in a $\Delta$-dependent manner \cite{Botter2011,Fabre1994,Mancini1994}. At our detector, we expect sub-shot-noise squeezing of only a few percent, due to the interference, optical losses, and the heterodyne detector's quantum efficiency. For each run of the experiment, we extract the noise power spectral density (PSD) of the demodulated heterodyne signal.  To obtain sufficient sensitivity to observe low levels of squeezing, we average data from nearly 2000 cycles of the experiment, corresponding to a total of 10 seconds of integration time. 

\begin{figure}[p]
	\centering
	\includegraphics[width=.95\textwidth]{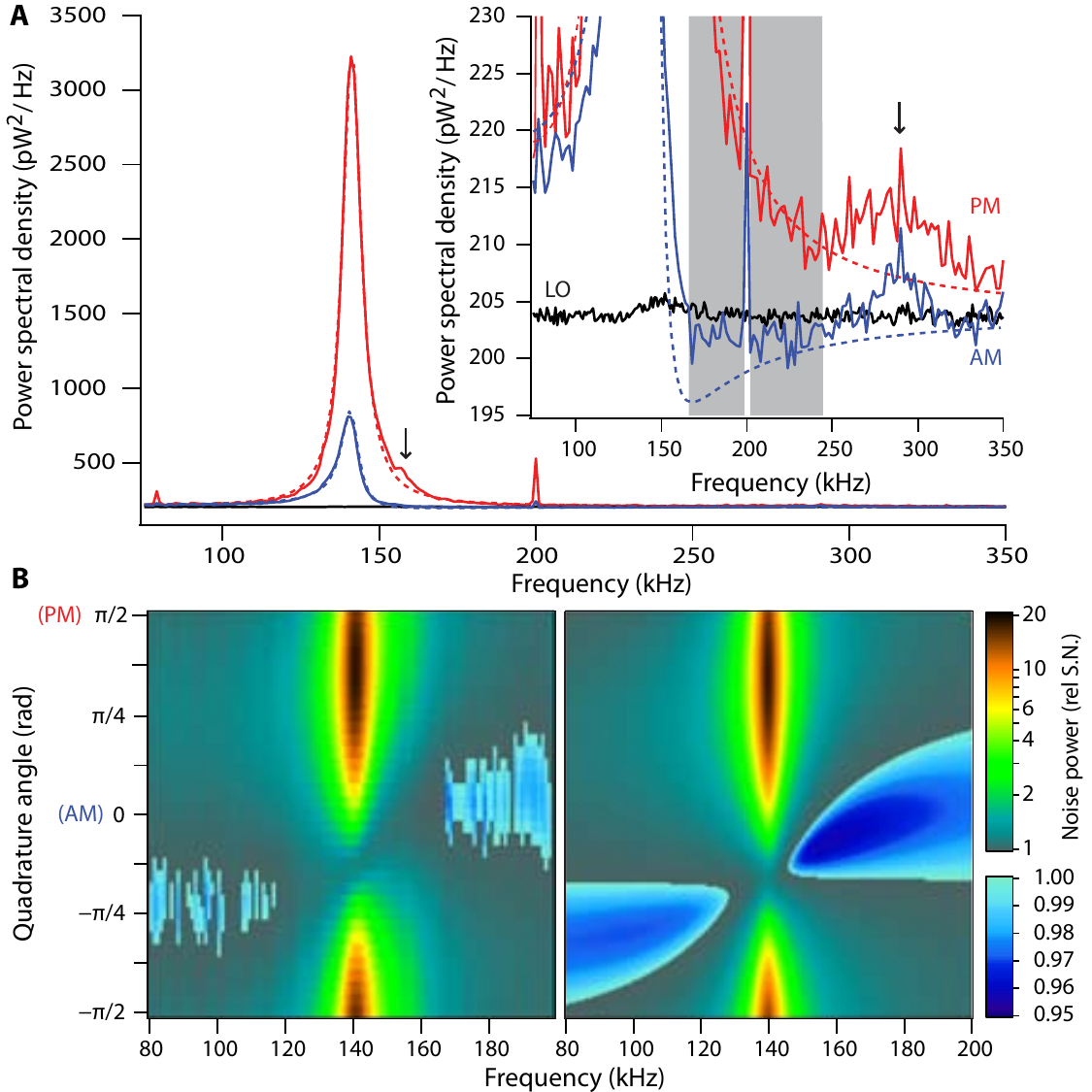}
	\dcaption{Power spectral densities (PSDs) measured with quantum noise driving the optomechanical system. \textbf{(A)} PSDs in the PM (red) and AM (blue) quadratures with corresponding theory (dashed lines). Theory parameters,  determined by fitting both the PM PSD and the technical noise measured in empty cavity spectra (shown in Supplemental Information), give the zero-parameter AM theory line. Shot noise (SN) is determined from the LO PSD (black).  A technical noise spike is visible at 200 kHz. \textbf{(Inset)} Expanding the vertical axis reveals the AM PSD to be below SN for frequencies above $\wm$. We average the AM PSD over regions (gray) where extra noise, from optomechanical response  (black arrows) not predicted by linear theory, is visible in the PM quadrature. \textbf{(B)} PSD versus quadrature, relative to SN. The color scale changes for squeezed regions. Measured PSD (left) suggests an additional region of squeezing below $\ws$. The theory (right) indicates this feature would be significant in the absence of technical noise.}
	\label{fig:Squeeze}
\end{figure}

Averaged over the frequency range indicated in Fig.~\ref{fig:Squeeze}, we observe the optomechanically transduced PSD in the AM quadrature to be ponderomotively squeezed to $99.0 \pm 0.1\%$ of the measured noise power of uncorrelated vacuum. The linear theory predicts AM squeezing to be 96.3\% of shot noise, as shown in Fig.~\ref{fig:Squeeze}.  
Squeezing can be masked by technical optical noise (Brownian noise at low frequencies and additional narrow-band spikes) or by response which deviates from the linear theory.  In order to identify frequencies where these effects occur without introducing bias in our analysis of the AM quadrature, we have used independent measurements from the orthogonal PM quadrature. We also see squeezing in a frequency range below $\ws$, at a quadrature angle around $-40^\circ$ from AM, matching the prediction of the extra-cavity linear theory \cite{Botter2011}, both shown in Fig~\ref{fig:Squeeze}b. At this quadrature, in regions without technical noise, squeezing is found to be $99.6\% \pm 0.2\%$ of shot noise. 

In the measured PSD, we find significant deviations from the linear theory. From the PM quadrature, we identify two regions of unexpected additional response, one near $\wm$, and one near $\wm + \ws$. Given that the single-photon strong coupling parameter \cite{Nunnenkamp2011,Rabl2011} in our system is $g/\wm > 0.4$, the latter peak is likely due to the nonlinearities inherent in the optomechanical interaction. Nonlinear wave-mixing between the Brownian noise at lower frequencies with the amplified response near $\ws$ may also explain why the observed squeezing is less than expected. Nonlinear effects due to single-photon strong-coupling will be the subject of future investigation. The noise peak at $\wm$ may be due to the disturbance of the center-of-mass mode by the remaining axial mechanical modes of the atomic ensemble, all of which occur at frequencies near $\wm$.  These additional modes, which are not sensed directly by the cavity field, serve as the mechanical thermal bath whose energy exchange with the center-of-mass mode contributes to $\Gm$.

Our observation of ponderomotive squeezing was made possible by constructing a system that is significantly driven by the radiation-pressure fluctuations due to shot noise (RPSN). Previous experiments quantified RPSN bolometrically \cite{Murch2008} whereas here we detect the optomechanically transduced shot noise directly.  We confirm from the  empty-cavity heterodyne record (shown in Supplemental Information), that, while Brownian technical noise contributes significantly to the intracavity fluctuations for $\omega < 2\pi \times 75$~kHz, the noise power spectral density is dominated by shot noise at higher frequencies. We quantify the contribution of RPSN to the PSD shown in Fig.~\ref{fig:Squeeze} by comparing the theoretical response to quantum fluctuations to the total measured response. For instance, at $\ws = 2\pi \times 140$~kHz, 97\% of the PM PSD is expected from RPSN alone, whereas the remaining 3\% may result from residual technical noise. This observation, in combination with the observed squeezing, demonstrates that we have observed both the amplification and suppression of the vacuum field due to its interaction with the motion of a mechanical element.

In this work, the observation of RPSN and ponderomotive squeezing is enabled by two key features. First, the mechanical element is optically trapped, decoupling it from the surrounding environment. Additionally, we operate deep in the unresolved-sideband regime, where $\wm \ll \kappa$. In this limit, the sensitivity of optical fluctuations to thermal occupation of the mechanical oscillator --- which plagued previous studies \cite{Verlot2010,Marino2010, Teufel2011,Weis2010,Safavi-Naeini2011a} ---  is significantly reduced and can be neglected. Understanding ponderomotive squeezing is key to reducing noise in optomechanical force detectors, e.g. gravitational wave observatories \cite{Corbitt2006}. Our results show theoretical treatment beyond the linear model will be necessary for a full understanding of ponderomotive squeezing. In addition, how signals from external forces are affected by the optomechanical amplification and attenuation needs further experimental and theoretical exploration.

\bibliographystyle{Science}
\bibliography{PonSqueezeExp}

\newpage
\section{Supplemental Information}
\setcounter{equation}{0}
\setcounter{figure}{0}
\setcounter{page}{1}
\renewcommand{\theequation}{S\arabic{equation}}
\renewcommand{\thefigure}{S\arabic{figure}}
\renewcommand{\thepage}{\hspace{-1.75in}\roman{page}}

\subsection{Experiment Methods}

Beginning a magnetic trap produced by the microfabricated atom-chip, the atoms are transferred into and trapped by a one-dimensional optical lattice formed by 850 nm light that resonates within a TEM$_{00}$ mode of the optical cavity. Adjusting the 850 nm light's intensity, the oscillation frequency is set to $\wm=2\pi\times155.5$~kHz, as measured via parametric heating of the atomic ensemble. The cavity optomechanical system is then probed with circularly polarized 780 nm light detuned by $-2\pi\times31$~GHz from the D2 transition of $^{87}$Rb, and frequency stabilized at a detuning of $\Dpc =- 2 \pi \times 1.0$~MHz from a second TEM$_{00}$ mode. 

To observe the optomechanical response in our system, we detect the signal beam's transmission through the cavity with a balanced heterodyne receiver. As shown in Fig.~\ref{fig:Schematic}, before the cavity we split the probe laser into two paths. A small fraction, the signal beam, is frequency shifted by $\fb=10$ MHz prior to entering the cavity. The rest bypasses the cavity to serve as the local oscillator (LO). The signal beam is intensity-stabilized before the cavity. Using the method described in previous work \cite{Purdy2010}, we monitor the probe intensity in transmission by sending 10\% of the signal exiting the cavity to a single-photon counting module (SPCM). The SPCM signal is used in a low-bandwidth lock of the signal beam's detuning $\Dpc$  from the shifted cavity resonance. The SPCM signal feeds back via an acousto-optic modulator (AO) to the probe frequency; by maintaining an average photon occupation $\nbar$ inside the cavity, $\Dpc$ is also fixed.  The rest of the cavity output is sent through an optical isolator and then recombined with the LO on a polarized beam splitter (PBS). Mode-matching is performed via coupling through a single-mode fiber. The fiber output is evenly split on a second PBS and both paths are sent to a 80 MHz New Focus 1807 balanced photoreceiver. A voltage-controlled liquid crystal polarizer is used to maintain equal power on both photodiodes to $\sim1\%$.

The photoreceiver output is sent through a bias-T and a 19.7~MHz low-pass filter before being amplified and recorded at $80\times10^6$ samples per second on a digital oscilloscope. This signal in turn is digitally demodulated at the intermediate frequency, $\fb$, to obtain conjugate quadratures, I and Q. As a result of employing heterodyne detection, the relative path lengths of the LO and signal beam need not be stabilized. We correct for variations of the path length by identifying the phase drift using a 10 Hz low-pass digital boxcar filter, and then applying a digital rotation to I and Q to extract the AM and PM signals.

We verify the quantum-limited nature of the LO near $\fb$ by ensuring that its variance scales linearly with input power. The variance-to-power relationship was found to be linear with 1.7\% uncertainty for powers ranging from 300 $\mu$W to 2.5 mW over a bandwidth of 2 MHz centered on $\fb$. The LO power was intensity stabilized to a value near 980 $\mu$W, a level well within the linear range of the photoreceiver. Digital rotation is not performed for the LO-vacuum detections as there is no meaningful low-frequency drift.

The detection efficiency is calibrated by measuring the optical spring effect at various light levels to determine a ratio of counts-per-second on the SPCM to $\nbar$. A technical signal is then measured relative to shot noise on both detectors in addition to measuring the optical power to the detector. We find the SPCM to have a detection efficiency of $\varepsilon=1.4\%$ and the heterodyne photoreceiver to have a detection efficiency of $\varepsilon=10.1\%$
\subsection{Linear Optomechanical Theory}
As discussed in detail in reference \cite{Botter2011}, input fluctuations $\alfop(\omega)$ can be re-expressed after being filtered by the cavity:
\begin{equation}
\bfop(\omega) =\sqrt{2\kappa} \frac{\kappa + i (\Delta - \omega)}{(\kappa - i \omega)^2 + \Delta^2}\alfop(\omega)
\end{equation}
The intracavity field can therefore be represented by Equation \ref{Ha} in the text. Assuming a high Q, that is $\wm / \Gm \gg 1 $, the optomechanical gain (OMG) is:
\begin{gather}
\label{Gwt} G(\omega) = \frac{\left(i \kappa - \Dpc + \omega \right)\sopt(\omega)}{\wm^2 +\Dpc \sopt(\omega) -\omega^2-i\omega \left(\Gm + \Go \right)}\\
\label{OMp} \sopt(\omega)=\frac{4m\gom^{2}\,\bar{n} \omega_{m}}{\kappa^{2}+\Dpc^{2}-\omega^{2}}
\qquad \qquad \Go(\omega)=\frac{2\kappa(\wm^{2}-\omega^{2})}{\kappa^{2}+\Dpc^{2}-\omega^{2}}
\end{gather}
In the unresolved sideband limit $\wm \ll \kappa$, the effect of the frequency dependence of $\sopt$ on the OMG becomes negligible. In this limit, the maximum intracavity squeezing attainable occurs at $\wm$ when $\Delta = \kappa$, suppressing the AM field by a factor of $\approx 2 \bar{n} \gom^2 / \Gm \kappa = C_\mathrm{opt}$, described in the literature as the collective optomechanical cooperativity.

\newcommand{\w}{\omega}
The field outside the cavity $\alfop_\mathrm{out}(\w)$ is related to the intracavity field via the output boundary condition \cite{walls1994},
\begin{equation}
\alfop_\mathrm{out} = \sqrt{2\kappa} \afop - \alfop_\mathrm{in}.
\end{equation}
A detection quadrature of the output field is defined by
\begin{equation}
X^{\theta}(\w) \equiv \alfop_\mathrm{out}(\w) e^{i\theta} + \alfop^\dagger_\mathrm{out}(-\w) e^{-i\theta}.
\end{equation}
Using the above definition, $X^0$ is identified with the AM quadrature and $X^{\pi/2}$ is identified with the PM quadrature.
In heterodyne detection, the power spectral density in a given quadrature is
\begin{equation}
S^\theta_\mathrm{het} = \frac{P_\mathrm{LO}\, \hbar \w_\mathrm{LO}}{2} \left [1 + \omega_\mathrm{BW} \langle X^\theta(-\w) X^\theta(\w) \rangle \right ]. 
\end{equation}
In the above, $P_\mathrm{LO}$ is the local oscillator power, $\w_\mathrm{LO}$ is the absolute carrier frequency of the LO, $ \omega_\mathrm{BW}$ is the bandwidth of the Fourier transform, and the factor of 1 arises from demodulation of LO shot noise at a frequency $\w_\mathrm{LO}-\w_p$, whereas the signal of interest arises from demodulation of the LO-probe beat at $\w_p-\w_\mathrm{LO}$, where $\w_p$ is the probe carrier frequency.
Finally, we note that cavity and optical losses lead to a detected PSD of
\begin{equation}
S^\theta_\mathrm{lossy} = \varepsilon S^\theta_\mathrm{het} + (1 - \varepsilon) P_\mathrm{LO}\, \hbar \w_\mathrm{LO},
\end{equation}
where $\varepsilon$ is the product of the probability $\varepsilon_\mathrm{cav}$ for light to exit the cavity via the output mirror, and the detection efficiency $\varepsilon_\mathrm{det}$.

\subsection{Response to an AM drive}
To characterize the response of the cavity-optomechanical system, a coherent modulation is added to the input light. Using a function generator and mixer we generate AM sidebands on the RF signal driving  an AO in the signal path after the LO has already been split off (see Fig.\ref{fig:Schematic}a). For each run, we vary the frequency at which the AM tone is applied and perform a superheterodyne detection. The photoreceiver signal is first demodulated at the intermediate frequency $\fb$ to generate the response in the AM and PM quadratures. We then demodulate each of these data traces a second time at the applied drive frequency to obtain the components in phase and out of phase with the drive. We normalize the optomechanically transduced signal components by the in-phase components of the signals measured through the cavity empty of atoms.

We then fit the normalized components to the linear theory to obtain $\ws$ and $\Gm$. The squared magnitude and argument of these signals, and the corresponding theory lines, are shown in Fig.~\ref{fig:Squash}. Shot noise, determined by averaging the PSD of the AM and PM quadratures while excluding the drive frequency, is rectified in calculating the squared magnitude. We subtract the measured shot-noise from the data to obtain the squared magnitude of the OMG for the drive tone. 

\begin{figure}
	\centering
	\includegraphics[]{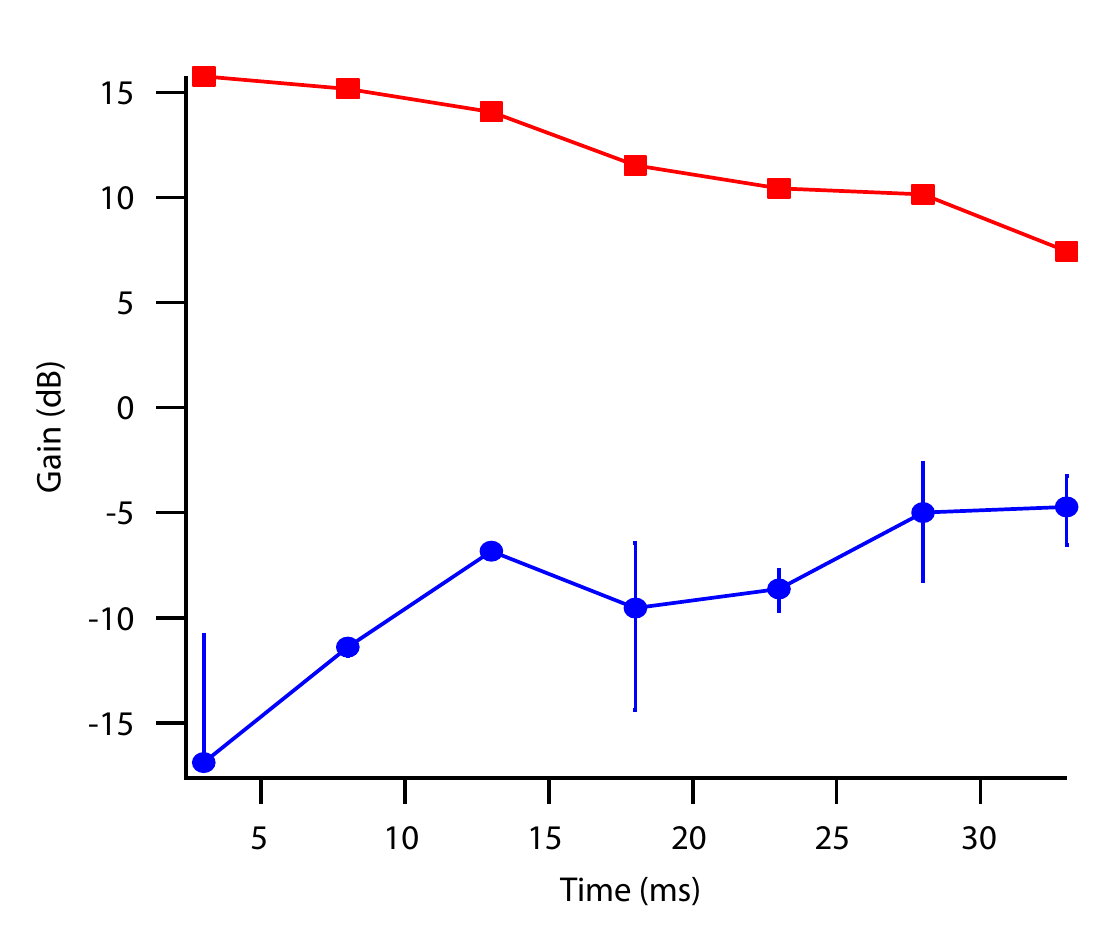}
	\dcaption{Temporal dependence of the optomechanical gain, as measured via the classical "network analysis" experiment.  Shown are the gains of the amplification and attenuation regions of the spectral response in the AM quadrature.  Each point is averaged over 5 ms.  Data for amplification are averaged over drives at 132.0 and 132.5 kHz; data for attenuation are averaged over drives at 155.0, 155.5, and 156.0 kHz.}
	\label{fig:gvt}
\end{figure}

Because the atoms are initially at a temperature below the limit achievable from sideband cooling \cite{Marquardt2007,Wilson-Rae2007}, fluctuations of the light field heat the atomic ensemble despite operating detuned on the ``cooling" side of cavity resonance \cite{Murch2008}. As a function of time spent probing the atomic ensemble, we observe that the system's optomechanical response is reduced.  Supplemental Fig.~\ref{fig:gvt} shows the reduction in gain for both the amplification maximum and squashing minimum as the atoms are probed for ever longer times.  Although the exact cause of this reduction is unknown, we expect that optomechanical heating causes an expansion of the atomic cloud over time, resulting in the atoms exploring the anharmonic character of their sinusoidal trapping potential, as well as reduced optomechanical coupling.  We observe gain consistent with our prediction from linear harmonic theory over the first 5 ms of each experimental run, and hence we restrict our data collection to this interval.

\subsection{Response to quantum fluctuations}
During the accumulation of the shot-noise driven data, we rejected one percent of the experiment cycles on the basis of abnormally high atom loss rates. These appeared to be correlated with transient spikes in the error signal of the trapping laser lock. The technical and shot noise were determined from the LO-empty cavity and LO-vacuum data records of every cycle not rejected. Before locking the probe to the side of the empty-cavity resonance, we sweep the probe across the cavity resonance and fit the transmission to determine the cavity frequency. On every shot we compare the probe frequency upon locking to the cavity resonance. Averaging over all the runs that contribute to the squeezing data, we measured the detuning to be $\Dpc/\kappa = -0.575 \pm .08$. To account for the frequency noise in $\Dpc$ we convolve the expected response with a Gaussian distribution of detunings centered on the mean. We fit the PM quadrature to this function to obtain $\ws = 2\pi \times 140.8$ kHz, $\Gm = 2\pi \times 3.2$ kHz, and a standard deviation of the spread in detunings of $0.14\, \kappa$. We use these values to calculate the theoretical response in the orthogonal AM quadrature.

\begin{figure}
	\centering
	\includegraphics[]{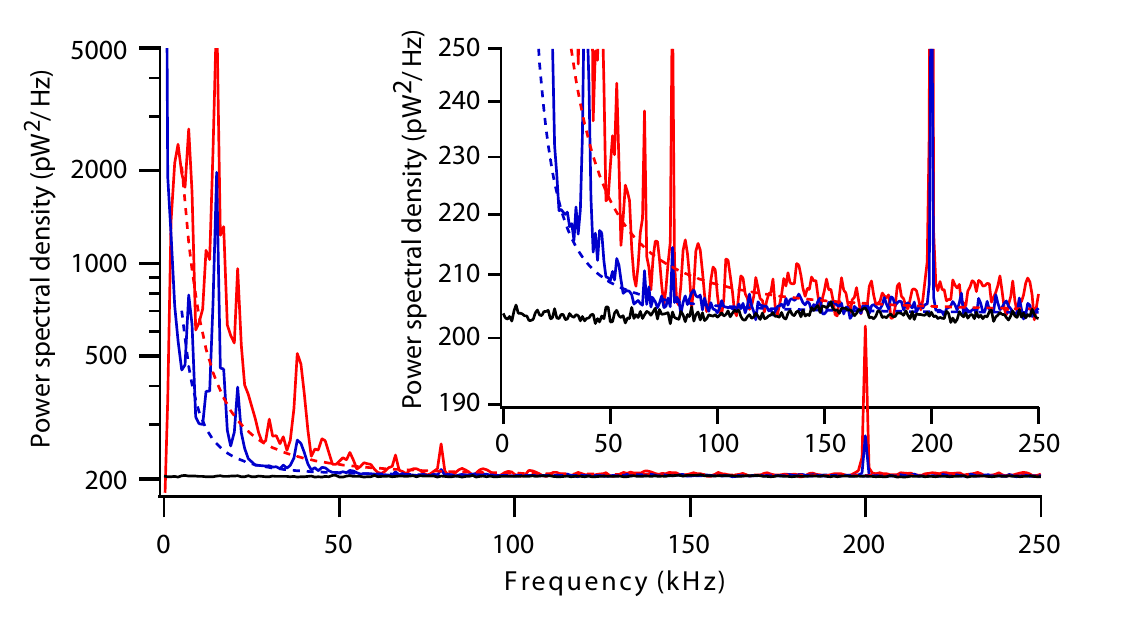}
	\dcaption{PSDs of the residual technical noise measured through the cavity empty of atoms. PSDs in the PM (red) and AM (blue) quadratures, have noise spikes on top of a Brownian noise background (dashed lines). \textbf{(Inset)} Expanding the vertical axis reveals technical noise to be only a few percent of shot noise (black) for frequencies above $75$~kHz. The technical noise is found mostly in the PM quadrature, whereas only AM noise contributes to radiation pressure fluctuations.}
	\label{fig:ECplot}
\end{figure}

The PSD of the technical noise are show in Supplemental Fig.~\ref{fig:ECplot}. We fit, excluding narrow-band peaks, to find the Brownian noise background to be $460/\nu^2$~$\mathrm{\mu W^2\, Hz}$ and $120/\nu^2$~$\mathrm{\mu W^2 \,Hz}$ in the PM and AM quadratures, respectively, where $\nu$ is the frequency. The noise is mostly phase noise, which does not contribute to radiation pressure fluctuations. Thus we expect only a minimal contribution in the optomechanically transduced spectra from the residual AM technical noise at frequencies greater than $75$~kHz.

In the shaded region of Fig.~\ref{fig:Squeeze}, the average PSD for the LO-vacuum beat was measured to be $203.78 \pm 0.05$~$\mathrm{pW^2/Hz}$ whereas the average AM PSD for the beat of LO with the ponderomotively affected cavity field was $201.8 \pm 0.2$~$\mathrm{pW^2/Hz}$. The detector-contributed noise power of $0.9$~$\mathrm{pW^2/Hz}$ is negligible and does not affect the finding of the optomechanically transduced noise to be ponderomotively squeezed to $99.0 \pm 0.1\%$ of the measured shot noise.
\end{document}